\documentclass[12pt]{article}
\usepackage{epsfig}
\usepackage{amssymb,amsmath}

\newcommand{\bea}{\begin{eqnarray}}
\newcommand{\eea}{\end{eqnarray}}

 \makeatletter
\def\alt{\mathrel{\mathpalette\gl@align<}}
\def\agt{\mathrel{\mathpalette\gl@align>}}
\def\gl@align#1#2{\lower.6ex\vbox{\baselineskip\z@skip\lineskip\z@
\ialign{$\m@th#1\hfil##\hfil$\crcr#2\crcr\sim\crcr}}} \makeatother

\begin{document}
%\begin{flushright}
%KEK-TH-XXXX
%\end{flushright}
%
\vspace*{1.0cm}

\begin{center}
\baselineskip 20pt 
{\Large\bf 
Gauss-Bonnet braneworld cosmological effect \\
on relic density of dark matter 
}
\vspace{1cm}

{\large 
Nobuchika Okada  and  Satomi Okada
}
\vspace{.5cm}

{\baselineskip 20pt \it
Institute of Particle and Nuclear Studies, \\ 
High Energy Accelerator Research Organization (KEK),  \\
1-1 Oho, Tsukuba, Ibaraki 305-0801, Japan} 

\vspace{.5cm}

\vspace{1.5cm} {\bf Abstract}
\end{center}

In Gauss-Bonnet braneworld cosmology, the Friedmann equation 
 of our four-dimensional universe on 3-brane is modified 
 in a high energy regime (Gauss-Bonnet regime), 
 while the standard expansion law is reproduced in low energies 
 (standard regime). 
We investigate the Gauss-Bonnet braneworld cosmological effect 
 on the thermal relic density of cold dark matter 
 when the freeze-out of the dark matter occurs in the Gaugss-Bonnet regime. 
We find that the resultant relic density is considerably reduced 
 when the transition temperature, which connects the Gauss-Bonnet 
 regime with the standard regime, is low enough. 
This result is in sharp contrast with the result previously 
 obtained in the Randall-Sundrum braneworld cosmology, 
 where the relic density is enhanced. 

\thispagestyle{empty}

%\bigskip
\newpage

\addtocounter{page}{-1}

%%%%%%%%%%%%%%%%%%%%%%%%%%
%\baselineskip 36pt
% Main body
%%%%%%%%%%%%%%%%%%%%%%%%%%
\baselineskip 18pt
%%%%%%%%%%%%%%%%%%%%%%%%%%

%%%%%%%%%%%%%%%%%%%%%%%%
\section{Introduction} 
%%%%%%%%%%%%%%%%%%%%%%%%
Recent various cosmological observations, in particular, 
 the Wilkinson Microwave Anisotropy Probe (WMAP) 
 satellite \cite{WMAP}, have established 
 the $\Lambda$CDM cosmological model with a great accuracy 
 and the relic abundance of the cold dark matter is estimated as 
 (68 \% C.L. uncertainties) 
\begin{equation}
\Omega_{CDM} h^2 = 0.1131 \pm 0.0034 .
\end{equation}
However, to clarify the identity of the dark matter particle 
 is still a prime open problem in particle physics and cosmology. 
Since the Standard Model (SM) has no suitable candidate for 
 the cold dark matter, the observation of the dark matter  
 suggests new physics beyond the SM in which the dark matter  
 particle is naturally provided. 
Many candidates for dark matter have been proposed 
 in various new physics models, for example, 
 the neutralino in supersymmetric models is one of 
 the promising candidates \cite{DMreview}.

Among several possibilities, the dark matter as the thermal relic 
 may be the most plausible scenario 
 since in this case, the relic abundance of the dark matter 
 is insensitive to the history of the early universe 
 before the freeze-out time of the dark matter, 
 such as the mechanism of reheating after inflation etc. 
This scenario allows us to estimate the dark matter relic density 
 by solving the Boltzmann equation \cite{Kolb}, 
\bea 
\frac{d Y}{d x}
= -\frac{s\langle\sigma v\rangle}{x H}(Y^2-Y_{EQ}^2) ,
\label{Y;Boltzmann}
\eea 
where $Y=n/s$ is the yield defined by the ratio of  
 the dark matter number density $(n)$ to the entropy density 
 of the universe $ s = 0.439 g_* m^3/x^3$, 
 $g_* \sim 100$ is the effective total number of relativistic 
 degrees of freedom, and $x=m/T$ is the ratio between 
 the dark matter mass ($m$) and the temperature of the universe ($T$). 
The yield in equilibrium $Y_{EQ}$ is written as 
 $ Y_{EQ} = 0.145 (g/g_*) x^{3/2} e^{-x}$ for $x \gtrsim 3$ with $g$ 
 being the degrees of freedom of the dark matter. 
In the standard cosmology, the Hubble parameter is described 
 by the total energy density of the universe 
 $(\rho=(\pi^2/30) g_* (m/x)^4) $ 
 through the Friedmann equation of the form 
\bea
 H^2 = \frac{8\pi}{3 M_{Pl}^2} \rho,    
\label{FriedmannEq}
\eea 
 where $M_{Pl}=1.22 \times 10^{19}$ GeV is the Planck mass. 
Using explicit formulas presented above, the Boltzmann equation 
 can be rewritten into the form 
\bea 
\frac{d Y}{d x}
= -\frac{\lambda}{x^2} \langle\sigma v\rangle (Y^2-Y_{EQ}^2) 
\label{Y;Boltzmann2}
\eea 
with an $x$-independent constant 
 $\lambda = xs/H = 0.26 \sqrt{g_*} m M_{Pl}$.

Solving the Boltzmann equation, we can obtain the thermal relic 
 abundance of the dark matter at the present universe. 
As is well known, an approximate formula of the solution 
 to the Boltzmann equation (for the S-wave annihilation process) 
 can be described as 
\bea 
Y(\infty) \simeq 
\frac{x_d}{\lambda \langle \sigma v \rangle},  
\eea 
where $x_d = m/T_d$ with the decoupling temperature $T_d$. 
It is useful to express the relic density in terms 
of the ratio of the dark matter density to the critical density, 
$\Omega h^2 = m s_0 Y(\infty) h^2/\rho_c$, 
where $\rho_c = 1.1 \times 10^{-5} h^2$ cm$^{-3}$, 
$h = 0.71^{+0.04}_{-0.03}$ and $s_0 = 2900$ cm$^{-3}$ \cite{Kolb}: 
\bea 
 \Omega h^2 \simeq \frac{1.07 \times 10^9 x_d \; {\rm GeV}^{-1}}
 {\sqrt{g_*}M_{Pl}\langle\sigma {\rm v} \rangle}. 
\label{Omega} 
\eea 
In a typical dark matter scenario such as the WIMP 
(weakly interacting massive particle) scenario, 
$x_d \sim 23$ and thus, the dark matter relic density 
is controlled only by its annihilation cross section 
$ \langle \sigma v \rangle$.

Recently, the braneworld models have attracted lots 
 of attention as a novel higher dimensional theory. 
In these models, it is assumed that the standard model particles
 are confined on a ``3-brane'' while gravity resides in 
 the whole higher dimensional spacetime. 
 The braneworld cosmology based on the model first 
 proposed by Randall and Sundrum (RS) \cite{RS}, 
 the so-called RS II model, has been intensively 
 investigated \cite{braneworld}. 
It has been found that \cite{RSsolution} the Friedmann equation 
 in the RS cosmology leads to a non-standard expansion low 
 in high energies, while the standard cosmology is 
 reproduced in low energies.

Since the Hubble parameter is involved in the Boltzmann equation, 
 the dark matter relic abundance depends on the expansion law 
 of the early universe at freeze-out time. 
Therefore, if our universe is higher dimensional and obeys 
 the non-standard expansion law at the freeze-out time, 
 the resultant relic abundance of the dark matter will be altered. 
The RS braneworld cosmological effect on the dark matter 
 physics has been investigated in detail \cite{RSDM1, RSDM2, RSDM3, RSDM4}, 
 in particular, it has been shown that the resultant dark matter 
 relic abundance is considerably enhanced. 
In the same context of the RS braneworld cosmology, 
 other cosmological issues such as leptogenesis \cite{RSleptogenesis} 
 and the cosmological gravitino problem in supersymmetric models 
 \cite{RSgravitino} have also been examined.

In this letter, we investigate the non-standard cosmological effect 
on the dark matter relic abundance in the context of 
the Gauss-Bonnet (GB) braneworld cosmology. 
Once the Gauss-Bonnet term is added in the RS braneworld model, 
the Friedmann equation in high energies is modified from 
the one in the RS braneworld cosmology. 
We will show that the GB braneworld cosmological effect 
works to reduce the thermal relic abundance of the dark matter. 
This is in sharp contrast with the RS braneworld cosmological effect.

%%%%%%%%%%%%%%%%%%%%%%%%%%%%%%%%%%%%%%%%%%%%%%%%%%%%%%%%%%%
\section{Relic density under the modified Friedmann equation}
%%%%%%%%%%%%%%%%%%%%%%%%%%%%%%%%%%%%%%%%%%%%%%%%%%%%%%%%%%%
Let us begin with parameterizing a modified Friedmann equation 
 in a certain class of braneworld cosmological models such as 
\bea
 H = H_{st}  F(x_t/x), 
\label{parametrize1}
\eea
where $H_{st}$ is the Hubble parameter in the standard cosmology, 
and the function $F$ denotes the modification of the Friedmann 
equation in braneworld models. 
Phenomenologically viable models should reproduce 
 the standard cosmology in low energies, so that 
 we impose a condition: $F(x_t/x)=1$ for $ x_t/x \leq 1 $. 
Here, $x_t = T_t/m$ and we call $T_t$ ``transition temperature'' 
 at which the modified expansion law shifts into 
 the standard cosmological one. 
The model-independent constraints on $T_t$ is given 
 by the success of the Bing Bang Nucleosynthesis (BBN) 
 in the standard cosmology and the transition should 
 complete before the BBN era, $T_t \gtrsim 1$ MeV\footnote{
The precision measurements of the gravitational law 
 in sub-millimeter range lead to much more stringent constraint, 
 for example, $T_t \gtrsim 1$ TeV for the RS braneworld model \cite{RS}. 
However, this constraint is, in general, quite model-dependent 
 and can be moderated in some extended models \cite{maedawands}. 
In this paper, we consider only the model-independent BBN constraint 
 on the transition temperature.}. 
For concreteness, we assume
\bea
  F (x_t/x) =   \left( \frac{x_t}{x} \right)^\gamma,    
\label{parametrize2}
\eea
for $x_t/x > 1$ with a constant $\gamma $. 
We can see that this parameterization is in fact a good approximation 
 for Friedmann equations obtained in known braneworld cosmological models. 
For example, $\gamma= 2$ corresponds to the RS braneworld cosmology, 
while $\gamma= -2/3$ to the GB braneworld cosmology as we will see later. 
Here we keep $\gamma$ as a free parameter 
to make our discussion applicable to general braneworld  models 
(if such models exist).

First we approximately solve the Boltzmann equation 
 with the modified Friedmann equation in the non-standard cosmology. 
Eq.~(\ref{Y;Boltzmann2}) is modified into 
\bea 
\frac{d Y}{d x}
= -\frac{\lambda}{x^2} 
 \left( \frac{\langle \sigma v\rangle}{F(x_t/x)} \right)
 (Y^2-Y_{EQ}^2) ,
\label{Y;Boltzmann3}
\eea 
from which we can understand that the effect of 
 the modified Friedmann equation is equivalent 
 to modify the annihilation cross section in the standard cosmology; 
\bea 
 \langle \sigma v\rangle \to 
 \left( \frac{\langle \sigma v\rangle}{F(x_t/x)} \right)  
 = \langle \sigma v \rangle 
   \left( \frac{x}{x_t}\right)^\gamma 
\eea   
in the era $ x_t/x > 1$ which we are interested in. 
Eq.~(\ref{Omega}) implies that the braneworld cosmological effect 
 enhances (reduces) the thermal relic abundance 
 of dark matter for $\gamma > 0$ ($\gamma <0$).

For simplicity, we parameterize the thermal average of 
the annihilation cross section times the relative velocity 
as $\langle \sigma v \rangle = \sigma_n x^{-n}$  
with a (mass dimension 2) constant $\sigma_n$ 
and an integer $n$ ($n=0$ and 1 correspond to 
S-wave and P-wave processes, respectively). 
At the early time, the dark matter particle is 
in the thermal equilibrium and $Y$ tracks $Y_{EQ}$ closely.
To begin, consider the small deviation from the thermal distribution
 $\Delta =Y-Y_{EQ} \ll Y_{EQ}$.
The Boltzmann equation leads to 
\begin{eqnarray}
\Delta \simeq -\frac{ x^2 (x_t/x)^\gamma \frac{d Y_{EQ}}{d x}}
  {\lambda \sigma_n x^{-n} (2Y_{EQ}+\Delta )}
\simeq \frac{x_t^\gamma}{2\lambda \sigma_n } x^{2+n-\gamma},
\end{eqnarray}
where we have used an approximation formula 
 $d Y_{EQ}/dx \simeq - Y_{EQ}$. 
As the temperature decreases or equivalently $x$ becomes large,
 the deviation grows since $Y_{EQ}$ is exponentially dumping.
Eventually the decoupling occurs at $x_d$
 roughly evaluated as $\Delta(x_d) \simeq Y(x_d) \simeq Y_{EQ}(x_d)$. 
At further low temperature, $\Delta \simeq Y \gg Y_{EQ}$ is satisfied
 and $Y_{EQ}^2$ term in the Boltzmann equation can be neglected so that 
\begin{equation}
\frac{d \Delta}{d x} = 
 - \frac{\lambda \sigma_n }{x_t^\gamma} 
 x^{\gamma -n - 2} \Delta^2, 
\end{equation}
and the solution is formally given by 
\begin{eqnarray}
\frac{1}{\Delta(x)}= 
 \frac{1}{\Delta(x_d)} + 
 \frac{\lambda \sigma_n}{ (\gamma - n -1 ) x_t^\gamma} 
 \left( x^{\gamma -n -1} - x_d^{\gamma -n -1}   \right).  
\label{sol}
\end{eqnarray}

For the S-wave process ($n=0$), for example, 
 the well-known result in the standard cosmology 
 (corresponding to $\gamma=0$), 
 $1/Y(\infty) \simeq \lambda \sigma_0/x_d $, is obtained. 
When we take $\gamma=2$, our analysis here is the same as 
the one in \cite{RSDM1} for the RS braneworld cosmology. 
For $n=0$, for example, $\Delta(x)^{-1}$ is continuously growing 
 and this growth stops at $x = x_t$, so that 
 the resultant relic abundance has been found to be 
 $1/Y(\infty) \simeq \lambda \sigma_0 x_t$.  
Then, we obtain the ratio of the energy density of the dark matter
 in the RS braneworld cosmology ($\Omega_{(RS)}$)
 to the one in the standard cosmology ($\Omega_{(s)}$) such that
\bea 
\frac{\Omega_{(RS)}}{\Omega_{(s)}}
\simeq \left(\frac{x_t}{x_{d(s)}}\right), 
\eea
where $x_{d(s)}$ is the decoupling temperature
 in the standard cosmology. 
The relic abundance is enhanced by the RS braneworld effect 
 for $x_t > x_{d(s)}$, while it should saturate to the standard result 
 for $x_t < x_{d(s)}$.

On the other hand, if $\gamma < 0$, we arrived 
at the result which is in sharp contrast with 
the one in the RS braneworld cosmology. 
For simplicity, we consider only $n=0$ in the following 
 and the analysis for $n > 0$ is straightforward. 
In this case, we obtain 
\bea  
\frac{1}{Y(\infty)} \simeq \frac{\lambda \sigma_0}{1-\gamma} 
 \left( \frac{x_t}{x_d} \right)^{-\gamma} x_d^{-1},   
\eea 
and the ratio of the energy density of the dark matter 
 in this braneworld cosmology ($\Omega_{(b)}$)
 to the one in the standard cosmology ($\Omega_{(s)}$) 
 is evaluated as 
\bea 
\frac{\Omega_{(b)}}{\Omega_{(s)}} 
\simeq 
 (1-\gamma) \left( \frac{x_d}{x_t}\right)^{-\gamma} 
            \left( \frac{x_d}{x_{d(s)}}\right). 
\label{ratio}
\eea
Thus, the resultant relic density is reduced 
 in the case for $\gamma <0$ and $x_d < x_t $. 
This is nothing but the case that happens in the GB braneworld cosmology.

%%%%%%%%%%%%%%%%%%%%%%%%%%%%%%%%%%%%%%%%%%%%%%%%%%%%%
\section{Gauss-Bonnet braneworld cosmological effect}
%%%%%%%%%%%%%%%%%%%%%%%%%%%%%%%%%%%%%%%%%%%%%%%%%%%%%

Motivated by string theory considerations, it is a natural 
 extention to add higher curvature terms to the bulk gravity 
 action in the RS braneworld model \cite{Kimetal}. 
Among possible terms, the Gauss-Bonnet invariant 
 is of particular interests in five dimensions, 
 since it is a unique nonlinear term in curvature 
 which yields second order gravitational field equations. 
The five-dimensional gravitational action with the GB invariant 
 is given by 
\bea
 {\cal S} &=& \frac{1}{2\kappa_5^2} \int d^5x 
 \sqrt{-g_5}
 \left[
 - 2 \Lambda_5+ {\cal R} + 
 \alpha \left( {\cal R}^2 -4 {\cal R}_{ab} {\cal R}^{ab} 
 + {\cal R}_{a b c d}{\cal R}^{a b c d} \right) \right] \nonumber \\
&-& \int_{brane} d^4x 
 \sqrt{-g_4} \left( \sigma + {\cal L}_{matter} \right), 
\label{GBaction}
\eea
where $\kappa_5^2=8\pi/M_5^3$ with the five-dimensional 
Planck scale $M_5$, $ \sigma >0 $ is the brane tension, 
and $ \Lambda_5 <0 $ is the bulk cosmological constant. 
The RS model is recovered in the limit $\alpha \to 0$.

Imposing a $Z_2$ parity across the brane in an anti-de Sitter bulk 
and modeling the matters on the brane as a perfect fluid, 
the modified Friedmann equation on the spatially flat brane 
has been found as \cite{GBsolution1} \cite{GBsolution2} 
\bea 
 \kappa_5^2(\rho + m_\sigma^4) = 
  2 \mu \sqrt{1+\frac{H^2}{\mu^2}}
  \left( 3 - \beta +2 \beta \frac{H^2}{\mu^2} \right) ,  
\label{GBFriedmann1} 
\eea
where $\beta = 4 \alpha \mu^2 = 1-\sqrt{1 + 4 \alpha \Lambda_5/3}$ 
and $m_\sigma = \sigma^{1/4}$. 
There are four free parameters, 
 $\kappa_5$, $m_\sigma$, $\mu$ and $\beta$, 
 corresponding to the original free parameters, 
 $\kappa_5$, $\sigma$, $\Lambda_5$ and $\alpha$, 
 which are constrained by phenomenological requirements. 
To reproduce the Friedmann equation of the standard cosmology 
 with zero cosmological constant in the limit $H^2/\mu^2 \ll 1$,  
 we find two relations among the parameters: 
\bea 
 && \kappa_5^2 m_\sigma^4 = 2 \mu (3-\beta), \nonumber \\
 && \kappa_4^2 =\frac{8 \pi}{M_{Pl}^2} 
  = \frac{\mu}{1+\beta} \kappa_5^2 .
\label{relations} 
\eea
The modified Friedmann equation can be rewritten 
in the useful form \cite{GBsolution3} 
\bea
&& H^2 = \frac{\mu^2}{\beta} 
 \left[ (1- \beta) \cosh \left(\frac{2 \chi}{3} \right) -1 \right], 
  \nonumber \\ 
&& \rho + m_\sigma^4 = m_\alpha^4 \sinh \chi , 
\label{GBFriedmann2} 
\eea
where $\chi$ is a dimensionless measure of the energy density and 
\bea
 m_\alpha^4 = \sqrt{ 
 \frac{8 \mu^2 (1-\beta)^3}{\beta \kappa_5^4}} 
 = 2 \frac{\mu^2}{\kappa_4^2} 
  \sqrt{2 \frac{(1-\beta)^3}{\beta (1+\beta)^2}}. 
\eea
Here we have used Eq.~(\ref{relations}) to eliminate $\kappa_5$ 
in the last equality. 
In the same way, we express $m_\sigma$ as 
\bea
 m_\sigma^4 = 
 2 \frac{\mu^2}{\kappa_4^2} 
 \left( \frac{3-\beta}{1+\beta} \right). 
\eea

The evolution of the GB braneworld cosmology is characterized 
by the two mass scales, $m_\alpha$ and $m_\sigma$. 
Expanding Eq.~(\ref{GBFriedmann2}) with respect to $\chi$, 
we find three regimes for $m_\alpha > m_\sigma$: 
The GB regime for $\rho \gg m_\alpha^4$, 
\bea 
 H^2 \simeq 
 \left(  \frac{1+\beta}{4 \beta} \mu \kappa_4^2 \rho 
  \right)^{2/3},   
\eea
the RS regime for $m_\alpha^4 \gg \rho \gg m_\sigma^4$, 
\bea 
 H^2 \simeq 
  \frac{\kappa_4^2}{6 m_\sigma^4} \rho^2,  
\eea
and the standard regime for $m_\sigma^4 \gg \rho $, 
\bea 
 H^2 \simeq \frac{\kappa_4^2}{3} \rho.  
\eea
Since we are interested in the GB regime, 
let us simplify the evolution of the universe 
by imposing the condition $m_\alpha = m_\sigma$, 
which leads to 
\bea
 3 \beta^3 -12 \beta^2 + 15 \beta -2 = 0   
\eea
and hence, $\beta=0.151$. 
In this case, the RS regime is collapsed and 
there are only two regimes in the evolution of the universe. 
Applying the parameterization to the non-standard Friedmann 
equation in Eqs.~(\ref{parametrize1}) and (\ref{parametrize2}), 
\bea 
 H= H_{st} \left( \frac{\rho_t}{\rho}\right)^{1/6} 
  = H_{st} \left( \frac{x}{x_t} \right)^{2/3} 
\label{GBcase}
\eea 
 for $\rho > \rho_t$ or equivalently $ x < x_t$, 
 while $H=H_{st}$ for $\rho < \rho_t$, 
where
\bea
  \rho_t = \frac{27}{16} \left( \frac{1+\beta}{\beta}\right)^2
  \frac{\mu^2}{\kappa_4^2} \simeq 3.9 \mu^2 M_{Pl}^2 . 
\eea 
In Fig.~1, we show that our approximation for the Friedmann equation 
is in fact a good approximation to the exact form 
in Eq.~(\ref{GBFriedmann2}). 

%%%%%%%%%%%%%%%%%%%%%%%%%%%%%%%%%%%%%%%%%%%%%%%%%%%
\begin{figure}[ht]
\begin{center}
{\includegraphics*[width=.6\linewidth]{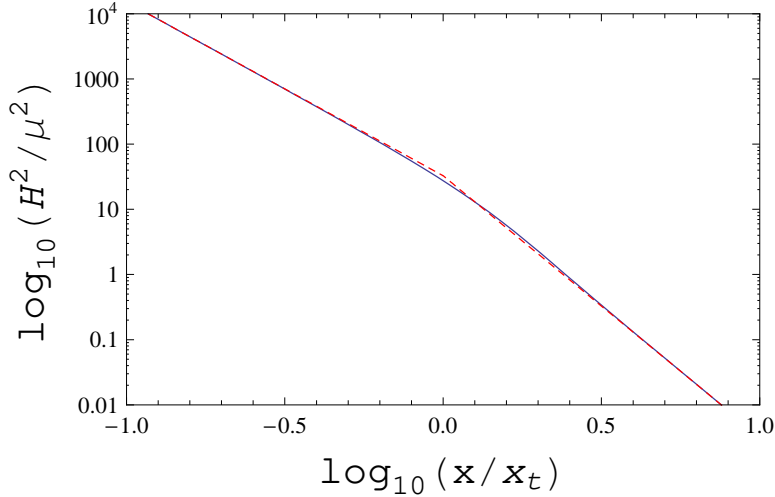}}
\caption{
The Hubble parameter as a function of $\rho$ in the GB braneworld cosmology. 
The solid line shows the exact formula of the Friedmann equation, 
 while the dashed (in red) corresponds to our approximation formula. 
Here, we have taken a unit $\kappa_4=1$. 
}
\end{center}
\end{figure}
%%%%%%%%%%%%%%%%%%%%%%%%%%%%%%%%%%%%%%%%%%%%%%%%%%% 

Now we are ready to see the GB braneworld cosmological effect 
on the dark matter  relic abundance. 
Eq.~(\ref{GBcase}) means that the modified Friedmann equation 
in the GB braneworld cosmology corresponds to $\gamma =-2/3$, 
and thus we find 
\bea 
\frac{\Omega_{(GB)}}{\Omega_{(s)}} 
\simeq 
  \frac{5}{3} 
   \left( \frac{x_t}{x_d}\right)^{2/3} 
   \left( \frac{x_d}{x_{d(s)}}\right)  
\label{ratio2}
\eea 
from Eq.~(\ref{ratio}). 
Therefore, the GB braneworld cosmological effect reduces the relic density 
and the reduction rate is controlled by the transition temperature. 
This is our main result. 
For the WIMP dark matter, the typical value of 
 the decoupling temperature is $x_d \sim 23$ and 
 this is not so much changing even under 
 the non-standard Friedmann equation. 
Thus, we expect 
\bea 
\frac{\Omega_{(GB)}}{\Omega_{(s)}} 
 \simeq \frac{5}{3} \left( \frac{x_t}{23}\right)^{2/3} . 
\eea 
For example, $\Omega_{(GB)}/\Omega_{(s)} \simeq 0.25 $ 
for $x_t=400$. 

%%%%%%%%%%%%%%%%%%%%%%%%%%%%%%%%%%%%%%%%%%%%%%%%%%%
\begin{figure}[ht]
\begin{center}
{\includegraphics*[width=.6\linewidth]{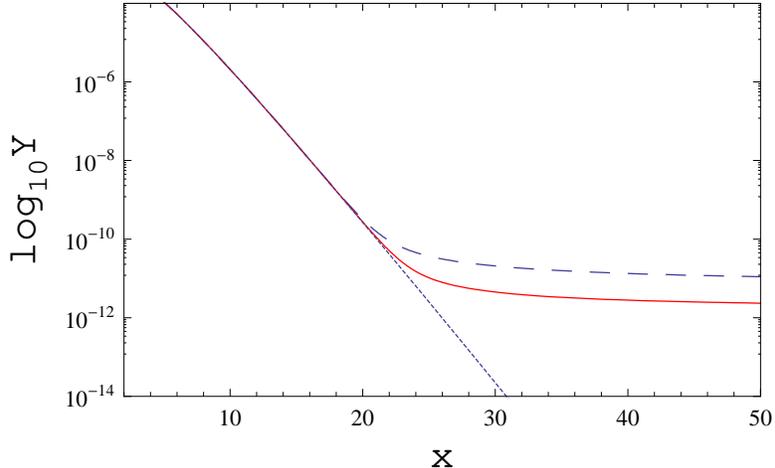}}
\caption{
Numerical solutions of the Boltzmann equation 
 in the Gauss-Bonnet braneworld cosmology (solid line) 
 and in the standard cosmology (dashed line). 
The dotted line corresponds to $Y_{EQ}$. 
The GB braneworld cosmological effect reduces 
 the resultant relic abundance from the one 
 in the standard cosmology. 
}
\end{center}
\end{figure}
%%%%%%%%%%%%%%%%%%%%%%%%%%%%%%%%%%%%%%%%%%%%%%%%%%%

Finally, let us check that our analytic result given above 
is a good approximation for the results 
from the numerical solution of the Boltzmann equation. 
Fixing the dark matter mass $m=100$ GeV, 
 $ \langle \sigma v \rangle = 10^{-5}/m^2 $ and $x_t=400$, 
 we numerically solve the Boltzmann equation 
 with the Friedman equations in the standard cosmology 
 and the GB braneworld cosmology, respectively. 
The numerical results are depicted in Fig.~2. 
Here, we obtain $\Omega_{(GB)}/\Omega_{(s)} \simeq 0.26$, 
 which is very close to our previous result from 
 analytic formulas.

%%%%%%%%%%%%%%%%%%%%%%%%%%%%%%%%%%%%%%%%%%%%%%%%%%%%%%%%
\section{Conclusions and discussions}
%%%%%%%%%%%%%%%%%%%%%%%%%%%%%%%%%%%%%%%%%%%%%%%%%%%%%%%%
We have investigated the thermal relic density
 of the cold dark matter in the braneworld cosmology, 
 in particular, the Gauss-Bonnet braneworld cosmology, 
 which is a natural extension of the Randall-Sundrum 
 braneworld model to include the higher curvature terms. 
We have modeled the modified Friedmann equation in such a way 
 applicable to general braneworld cosmological models 
 and analytically solved the Boltzmann equation 
 under some approximation. 
Applying this result to the Gauss-Bonnet braneworld cosmology, 
 we have found that the resultant relic density of the dark matter 
 is considerably reduced from the one in the standard cosmology, 
 when the freeze-out occurs well in the Gauss-Bonnet regime 
 in the evolution of the universe. 
This conclusion is in sharp contrast with the result 
 in the Randall-Sundrum braneworld cosmology, 
 where the relic density is enhanced 
 by the braneworld cosmological effect.

It is worth applying our results in this paper to concrete dark matter models 
 which have been investigated in the standard cosmology. 
For supersymmetric models with the neutralino dark matter, 
 the RS braneworld cosmological effect was analyzed \cite{RSDM2}. 
It has been shown that the allowed parameter region 
 for the neutralino dark matter consistent with the observed dark 
 matter density is dramatically modified from the one 
 in the standard cosmology and eventually disappears 
 as the transition temperature is lowered. 
This is because the RS braneworld cosmological effect enhances 
 the dark matter relic density while supersymmetric models, 
 in particular, the constrained minimal supersymmetric SM, 
 tend to predict an over-abundance of neutralino dark matter. 
On the other hand, the Gauss-Bonnet braneworld cosmological effect 
 reduces the relic density of the dark matter and therefore 
 can enlarge the allowed parameter region in supersymmetric models. 
A similar effect has been discussed in the scalar-tensor cosmology \cite{enlarge}. 
This enlargement of the cosmologically  allowed parameter region 
 has an impact on the sparticle search at the Large Hadron Collider (LHC). 
As have been investigated in \cite{RSleptogenesis, RSgravitino} 
 for the RS braneworld cosmology, 
 it would be interesting to consider other cosmological issues 
 also in the GB braneworld cosmology. 
We leave these subjects in future works.

%%%%%%%%%%%%%%%%%%%%%%%%%%%%%%%%%%%%%%%%%%%%%%%%%%%
%\begin{figure}[ht]
%\begin{center}
%{\includegraphics*[width=.6\linewidth]{Fig1.eps}}
%\caption{
%}
%\end{center}
%\end{figure}
%%%%%%%%%%%%%%%%%%%%%%%%%%%%%%%%%%%%%%%%%%%%%%%%%%%

%%%%%%%%%%%%%%%%%%%%%%%%%%%%%%%%%%
\section*{Acknowledgments}
%%%%%%%%%%%%%%%%%%%%%%%%%%%%%%%%%%
The authors would like to thank Shinji Komine for collaboration 
 in the early stage of this work. 
We are grateful to Osamu Seto for carefully reading 
 the manuscript and useful comments. 
We also wish to thank Andy Okada for his encouragement. 
This work of N.O. is supported in part by the Grant-in-Aid 
 for Scientific Research from the Ministry of Education, 
 Science and Culture of Japan, No.~18740170.

%%%%%%%%%%%%%%%%%%%%%%%%%%%%%%%%%

%%%%%%%%%%%%%%%%%%%%%
%  Figures
%%%%%%%%%%%%%%%%%%%%%
%%%%%%%%%%%%%%%%%%%%%%%%%%%%%%%%%%%%%%%%%%%%%%%%%
%\begin{figure}[t]\begin{center}
%\includegraphics[scale=1.2]{Fig1.eps}
%\caption{
%}
%\end{center}
%\end{figure}
%%%%%%%%%%%%%%%%%%%%%%%%%%%%%%%%%%%%%%%%%%%%%%%%%

\end{document}